\documentclass{article}
\usepackage{arxiv}

\usepackage[utf8]{inputenc} 
\usepackage[T1]{fontenc}    
\usepackage[hidelinks]{hyperref}       
\usepackage{url}            
\usepackage{booktabs}       
\usepackage[version=4]{mhchem}
\usepackage{amsfonts}       
\usepackage{mathtools}
\usepackage{nicefrac}       
\usepackage{microtype}      
\usepackage{lipsum}		
\usepackage{graphicx}
\usepackage[square,numbers,sort&compress]{natbib}
\usepackage{doi}
\usepackage{caption}
\usepackage{subcaption}
\usepackage{amsmath}
\usepackage{multirow}

\usepackage{bm}
\renewcommand{\vec}[1]{\bm{#1}}
   \newcommand{\mtx}[1]{\bm{#1}} 
\usepackage{url}
\usepackage[usenames,dvipsnames]{xcolor}
\usepackage[dvipsnames]{xcolor}
\definecolor{newcolor}{rgb}{.8,.349,.1}

\newcommand{\Rey}{{\mbox{\textit{Re}}}} 
\newcommand{\Reytau}{\mbox{\textit{Re}}_{\tau}}  
\newcommand{\Reyeff}{\mbox{\textit{Re}}_{\text{eff}}}  
\newcommand{\CFL}{{\mbox{\text{CFL}}}} 
   \usepackage{caption}
   \usepackage{subcaption}
   \usepackage{epsfig,epstopdf}
   \usepackage{ulem}
   
\title{On the performances of standard and kinetic energy preserving time-integration methods for incompressible-flow simulations}%

\author{ \href{https://orcid.org/0009-0009-5872-702X}{\includegraphics[scale=0.06]{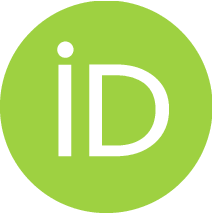}\hspace{1mm} Marco {Artiano}}\\
Institute of Mathematics\\
Johannes Gutenberg University Mainz\\
	Mainz, Germany \\
	\texttt{martiano@uni-mainz.de} \\
	\And
    \href{https://orcid.org/0000-0002-6518-3114}{\includegraphics[scale=0.06]{orcid.eps}\hspace{1mm} Carlo {De~Michele}}\\
	Dipartimento di Ingegneria Industriale\\
	Universit\`a di Napoli ``Federico II''\\
	Napoli, Italy \\
	\texttt{carlo.demichele2@unina.it} \\
	\And
    \href{https://orcid.org/0000-0003-0274-5260}{\includegraphics[scale=0.06]{orcid.eps}\hspace{1mm} Francesco {Capuano}}\\
    Department of Fluid Mechanics\\
    Universitat Politècnica de Catalunya $\cdot$ BarcelonaTech\\
	Barcelona, Spain \\
	\texttt{francesco.capuano@upc.edu} \\
	\And
	\href{https://orcid.org/0000-0003-4943-9551}{\includegraphics[scale=0.06]{orcid.eps}\hspace{1mm}Gennaro Coppola} \\
	Dipartimento di Ingegneria Industriale\\
	Universit\`a di Napoli ``Federico II''\\
	Napoli, Italy \\
	\texttt{gcoppola@unina.it} \\
}

\renewcommand{\shorttitle}{On the performances of standard and KEP time-integration methods}%

\hypersetup{
pdftitle={On the performances of standard and kinetic energy preserving time-integration methods for incompressible-flow simulations},
pdfsubject={physics.flu-dyn},
pdfauthor={Marco Artiano, Carlo De~Michele, Francesco Capuano, Gennaro Coppola}
pdfkeywords={large-eddy simulation, Runge-Kutta schemes, kinetic energy conservation, turbulent channel flow}
}

\begin{document}
\maketitle

\begin{abstract}
The effects of kinetic-energy preservation errors due to Runge-Kutta (RK) temporal integrators have been analyzed for the case of large-eddy simulations of incompressible turbulent channel flow. Simulations have been run using the open-source solver Xcompact3D with an implicit spectral vanishing viscosity model and a variety of temporal Runge-Kutta integrators. Explicit \textit{pseudo-symplectic} schemes, with improved energy preservation properties, have been compared to standard RK methods. The results show a marked decrease in the temporal error for higher-order pseudo-symplectic methods; on the other hand, an analysis of the energy spectra indicates that the dissipation introduced by the commonly used three-stage RK scheme can lead to significant distortion of the energy distribution within the inertial range. A cost-vs-accuracy analysis suggests that pseudo-symplectic schemes could be used to attain results comparable to traditional methods at a reduced computational cost. 
\end{abstract}

\section{Introduction} 

Guaranteeing the conservation of linear and quadratic invariants of the Navier-Stokes equations at a discrete level is considered to be of great importance for both direct and large-eddy simulations (LES) of turbulent flows~\citep{Coppola2019}. The lack of discrete kinetic energy preservation (KEP) can significantly alter the physical realism and robustness of scale-resolving simulations, for instance by: i) contaminating the energy cascade mechanism through artificial dissipation, or ii) causing spurious build-up of energy at the smallest wavenumbers that can eventually preclude the numerical stability of the computations~\citep{arakawa1981potential}. In incompressible flow models, kinetic energy is an inviscid invariant of quadratic type and also constitutes a norm of the solution; therefore, enforcing its discrete conservation can ensure unconditional stability without the addition of any type of artificial dissipation, an aspect which is especially desirable in LES where marginally-resolved scales possess a non-negligible amount of energy~\citep{kravchenko1997effect}.

In typical semi-discrete approaches, which are widely used in fluid dynamics, both spatial and temporal schemes contribute to the discrete kinetic energy balance. There are now a wide range of available approaches to build spatial discretization schemes with KEP, including high-order accuracy, and encompassing finite-difference, finite-volume, and finite-element methods and its derivations on both structured and unstructured meshes (see, for instance, recent review papers by \cite{perot2011discrete}, \cite{Coppola2019a}, \cite{veldman2021supraconservative} and references therein). 
On the other hand, much less attention has been devoted to the construction and application of kinetic-energy preserving time integration schemes.
Since the pioneering works from the Stanford group~\citep{rogallo1984numerical}, Runge-Kutta (RK) methods have become popular in the fluid dynamics community, and the majority of turbulence
simulations are nowadays performed using three-stage schemes (particularly the low-storage  scheme developed by~\cite{Wray1991}) in conjunction with fractional-step methods. However, research on the effect of time-integration errors in numerical simulations of turbulent flows has been relatively scarce in the existing literature~\citep{wang2013high}.

\cite{Moin} investigated the effect of large time steps for the turbulent channel flow, arguing that they might be responsible for unphysical behaviour on the turbulent structures or even the laminarization of the flow. Later, \cite{ham2002fully} developed a fully conservative (i.e. spatial + temporal) algorithm based on
the midpoint method, which proved to preserve energy exactly in time
in inviscid computations. \cite{verstappen2003symmetry} also applied the implicit midpoint rule to their symmetry-preserving
spatial discretization and recognized that the resulting
scheme is unconditionally stable on any mesh size and for any
time-step. More recently, \cite{sanderse2013energy} conducted a systematic
study of so-called \textit{symplectic} RK methods for the
incompressible Navier-Stokes equations. These methods are able to preserve  quadratic invariants, and therefore are also KEP. Results demonstrated the achievement of full KEP at the expense of using an implicit time integration approach. To reduce the cost of implicit time advancement while retaining good conservation properties, \cite{Capuano2017} proposed several \textit{pseudo-symplectic} methods, i.e. RK schemes of order $p$ that
preserve kinetic energy to order $q > p$. This class of schemes was later thoroughly assessed \citep{Capuano2019} for  LES of the Taylor-Green-Vortex. The results  emphasized the beneficial properties and efficiency of pseudo-symplectic RK schemes, while also warning on the potentially excessive dissipation introduced by classical three-stage methods. More recently, Ketcheson and coworkers introduced so-called \textit{relaxation} RK methods~\citep{ketcheson2019relaxation,ranocha2020relaxation}, that are able to preserve quadratic invariants to machine accuracy while retaining the accuracy and stability properties of the unmodified RK schemes. Although these methods were introduced in the context of compressible flow models, mostly to enhance entropy stability, they can also preserve---by construction---kinetic energy for incompressible flows, as shown by~\cite{CapuanoCoppola}. Despite these recent efforts, understanding the practical consequences of temporal dissipation remains, to date, elusive.

The aim of this work is to investigate the effects of time-integration errors (particularly the dissipative component) of standard and pseudo-symplectic RK schemes in the context of wall-bounded turbulent flows, at time steps close to the ones dictated by the linear stability limit. To this scope, the high-order finite-difference open-source flow solver Xcompact3D~\citep{Bartholomew2020} has been employed to run the simulations: the implicit LES (iLES) model is adopted through the use of a spectral vanishing viscosity operator, so that the extra-dissipation is enforced directly in the second derivative scheme of the diffusive term of the Navier-Stokes equations~\citep{Lamballais2011}. Classical and innovative Runge-Kutta (RK) time integrators have been implemented, in addition to the ones already present in the code.

The paper is organized as follows. Section~\ref{sec:mathematical_formulation} describes the mathematical formalism  and introduces the Runge-Kutta schemes employed in this work. Results are reported in Section~\ref{sec:results}, while concluding remarks are drawn in Section~\ref{sec:conclusions}.

\section{Mathematical formulation}\label{sec:mathematical_formulation}
The incompressible Navier-Stokes equations for a Newtonian fluid read:
\begin{equation}
\begin{aligned}
& \nabla \cdot \vec{V}=0 \\
& \frac{\partial \vec{V}}{\partial t}+(\vec{V} \cdot \nabla) \vec{V}=-\nabla p+\frac{1}{\Rey} \nabla^2 \vec{V},
\label{NSeq}
\end{aligned}
\end{equation}
where  $\vec{V}(\vec{x},t)$ is the (three-dimensional) velocity field, $p$ the pressure and $\Rey$ is the Reynolds number. 
In what follows we will assume that the system \eqref{NSeq} will be numerically integrated by using a semi-discretization procedure; i.e. the equations are firstly discretized in space and the resulting system of semi-discrete equations are advanced in time with a dedicated scheme. 
The procedure here investigated is agnostic with respect to the spatial discretization step, which can be performed  using finite difference (FD), finite volume (FV) or finite element methods, among others.
In general, spatial discretization leads to a semi-discrete formulation of the Eqs.~(\ref{NSeq}) which can be cast in the following form:
\begin{equation}
\begin{aligned}
& \mtx{M} \vec{u}(t)=\vec{r_1}(t) \\
&  \vec{\dot{u}}(t)=-\vec{C}(\vec{u}(t))\vec{u}(t)+\frac{1}{\Rey} \mtx{L} \mtx{u}(t)-\mtx{G} \vec{p}(t)+\vec{r_2}(\vec{u(t)}, t),
\label{DiscreteNS}
\end{aligned}
\end{equation}
where $\vec{u}(t)$ is the discrete velocity vector of size $N_{\vec{u}}$ gathering the three spatially discretized components of $\vec{V}$ along the Cartesian directions, $\vec{p}(t)$ is the analogous discrete pressure vector of size $N_{p}$, whereas $\vec{r_1}$ and $\vec{r_2}$ are the vectors resulting from the application of the boundary conditions in the continuity and momentum equations, respectively. 
The matrix $\mtx{C}(\vec{u})\in\mathbb{R}^{N_{\vec{u}}\times N_{\vec{u}}}$ is the convective operator that, acting on the $\vec{u}$ vector, returns the discrete (nonlinear) convective term, whereas 
the matrix operators $\mtx{M}\in\mathbb{R}^{N_p\times N_{\vec{u}}}$, $\mtx{L}\in\mathbb{R}^{N_{\vec{u}}\times N_{\vec{u}}}$ and $\mtx{G}\in\mathbb{R}^{N_{\vec{u}}\times N_p}$ are the discretization of the divergence, diffusion and gradient operators, respectively.
Their form depends on the details of the discretization procedure adopted and they can be considered constant matrices for stationary (non-moving) meshes. 

Equations~\eqref{DiscreteNS} constitute a system of Differential Algebraic Equations (DAE) and can be recast as a system of Ordinary Differential Equations (ODE) by enforcing the incompressibility constraint through the solution of a Pressure Poisson Equation (PPE). In this work, we adopt the approach detailed in \cite{Sanderse2012}, where the projection step is performed after the application of an explicit Runge-Kutta method to the unsteady part of the system in Eqs.~\eqref{DiscreteNS}. This procedure avoids the differentiation of the algebraic equations and allows the enforcement of the divergence-free constraint at each stage for non-stationary boundary conditions. 
For a general $s$-stage Runge-Kutta method one has
\begin{align}
    \vec{u}_i &= \vec{u}^n + \Delta t  \sum_{j=1}^{i-1} a_{ij} \widetilde{\boldsymbol{f}}_j
   \label{eq:RK_1}\\
    \vec{u}^{n+1} &= \vec{u}^n + \Delta t  \sum_{i=1}^{s} b_i \widetilde{\mtx{f}}_i \label{eq:RK_2}
\end{align}
where ${\widetilde{\vec{f}}}_j = \mtx{P}\vec{f}_j$, ${\vec{f}}_j = -\mtx{C}(\vec{u}_j) \vec{u}_j + \frac{1}{\Rey} \mtx{L} \vec{u}_j + \vec{r_2}(\vec{u}_j,t_j)$ and $\mtx{P} =  \mtx{I}- \mtx{G} \mtx{L}^{-1}\mtx{M}$ is the projection operator, satisfying $\mtx{M}\mtx{P}=\mtx{0}$.
The discrete counterpart of the continuous global kinetic energy equation can be derived starting from the Eqs.~\eqref{eq:RK_1}-\eqref{eq:RK_2} and reads for homogeneous and/or periodic boundary conditions \citep{Capuano2017}
\begin{equation}
    \frac{\Delta E}{\Delta t} = \frac{1}{\Rey}\sum_{i=1}^{s} b_j {\vec{u}_j}^T  \mtx{L}\mtx{u}_j - \frac{\Delta t}{2} \sum_{i,j=1}^{s} \left( b_i a_{ij} +  b_j a_{ji}-b_j b_i\right)\mtx{{\Tilde{f}_i}}^T\mtx{\Tilde{f}_j}\label{eq:DiscrKinEnergy_RK}
\end{equation}
where $\Delta E = E^{n+1} - E^n$ and $E$ is the discrete global kinetic energy.
In Eq.~\eqref{eq:DiscrKinEnergy_RK} the first term is a discrete analogue of the physical dissipation rate $\varepsilon_\nu$, whereas the second term is the temporal error $\varepsilon_{\text{RK}}$ associated with the Runge-Kutta procedure. 
To preserve the physical property that the variation of the kinetic energy for viscous flow is merely due to physical dissipation, the so-called M-condition $M_{ij} =  b_i a_{ij} +  b_j a_{ji}-b_j b_i = 0$ for $\forall i,j$ must be verified; symplectic RK schemes fulfill this requirement but they are necessarily implicit~\citep{Hairer2006}.
As in turbulent simulations explicit time integration procedures are usually preferred, due to the remarkable increase in computational efforts associated with implicit methods, 
in the present work we focus our attention on pseudo-symplectic schemes~(\cite{Aubry1998,Capuano2017}), which are explicit methods able to nullify the spurious production of energy up to a certain order of accuracy.

Classical RK schemes of $s$ stages are usually constructed to maximize the temporal order of accuracy $p$. In such case, $p$ also coincides with the pseudo-symplectic order $q$, defined so that the discrete evolution of the global kinetic energy for $\Rey \to \infty$ is 
 \begin{equation}
     \frac{\Delta E}{\Delta t} = O(\Delta t^q).
 \end{equation} 
A pseudo-symplectic method is one for which $q>p$. The schemes investigated in this analysis are 3p5q(4), 3p6q(5), 4p7q(6), where, by following the notation introduced in \cite{Capuano2017},  $n$p$m$q($s$) indicates a method with $p=n$ temporal order of accuracy, $q=m$ pseudo-symplectic order, and $s$ stages.
The coefficients of the schemes employed in this study are reported in Table~\ref{tab:schemes}.

\begin{table}[tb]
    \centering
    \begin{tabular}{c c c}
    \hline
          \hline $3p5q(4)$ & $3p6q(5)$ & $4p7q(6)$ \\
          \hline
 $a_{21}=3/8$   & $a_{21}=0.13502027922909$  & $a_{21}=0.23593376536652$ \\
 $a_{31}=11/12$ & $a_{31}=-0.47268213605237$ & $a_{31}=0.34750735658424$ \\
 $a_{32}=-2/3$  & $a_{32}=1.05980250415419$ & $a_{32}=-0.13561935398346$ \\
 $a_{41}=-1/12$ & $a_{41}=-1.21650460595689$ & $a_{41}=-0.20592852403227$ \\
 $a_{42}=11/6$  & $a_{42}=2.16217630216753$ & $a_{42}=1.89179076622108$ \\
 $a_{43}=-3/4$  & $a_{43}=-0.37234592426536$ & $a_{43}=-0.89775024478958$ \\
 $b_1=1/9$      & $a_{51}=0.33274443036387$ & $a_{51}=-0.09435493281455$ \\
 $b_2=8/9$      & $a_{52}=-0.20882668296587$ & $a_{52}=1.75617141223762$ \\
 $b_3=-2/9$     & $a_{53}=1.87865617737921$ & $a_{53}=-0.96707850476948$ \\
 $b_4=2/9$      & $a_{54}=-1.00257392477721$ & $a_{54}=0.06932825997989$ \\
                & $b_1=0.04113894457092$ & $a_{61}=0.14157883255197$ \\
                & $b_2=0.26732123194414$ & $a_{62}=-1.17039696277833$ \\
                & $b_3=0.86700906289955$ & $a_{63}=1.30579112376331$ \\
                & $b_4=-0.30547139552036$ & $a_{64}=-2.20354136855289$ \\
                & $b_5=0.13000215610576$ & $a_{65}=2.92656837501595$ \\
                &                            & $b_1=b_6=0.07078941627598$ \\
                &                            & $b_2=b_5=0.87808570611881$ \\
                &                            & $b_3=b_4=-0.44887512239479$ \\ 
\hline
    \end{tabular}
    \caption{Butcher tableau coefficients of pseudo-symplectic schemes \citep{Capuano2017}.}
    \label{tab:schemes}
\end{table}

 The dissipative effects of the temporal error can be efficiently investigated introducing the effective Reynolds number $\Reyeff$~\citep{Capuano2017}, that can be defined, starting from the discrete evolution of the global kinetic energy, as
\begin{equation}
    \frac{\Delta E}{\Delta t} = \frac{1}{\Rey} \Phi + \varepsilon_{\text{RK}} = \frac{1}{\Reyeff} {\Phi} 
    \label{DiscreteEnergyEquation}
\end{equation}
where the term  $\frac{1}{\Rey} \Phi$ represents the discretized physical dissipation rate, whereas $\varepsilon_{\text{RK}}$ is the temporal numerical dissipation. The second term indeed represents a source of error, due to the lack of the summation by parts rule of the RK integrator \citep{Capuano2018}. The expressions for the discrete counterpart of the physical dissipation rate, and for the RK temporal error read respectively (cf.~Eq.~\eqref{eq:DiscrKinEnergy_RK}) 
\begin{equation}
\varepsilon_\nu = \frac{1}{Re} \sum_{i = 1}^{s} b_j \vec{u}^T_j \mtx{L} \vec{u}_j
\end{equation}
\begin{equation}
  \varepsilon_{\text{RK}} =   - \frac{\Delta t}{2} \sum_{i,j=1}^{s} \left(b_i a_{ij} + b_j a_{ji}-b_j b_i\right) \mtx{{\Tilde{f}_i}}^T\mtx{\Tilde{f}_j}
\end{equation}
where $\vec{u}_j$ represents the $j$-th intermediate velocity field inside the $s$ inner stages of the RK procedure.

\section{Results}\label{sec:results}
 \begin{table}[tb!]
    \centering
    \begin{tabular}{ c c c c c c c }
\toprule
 $\textit{Re}_{\tau}$ & $L_x$ & $L_z$ & $Nx\times Ny \times Nz$ & $\Delta t$ & CFL & $\frac{\nu_0}{\nu}$\\ 
\midrule
 180 & $4\pi \delta $& $\pi \delta$ & $50 \times 33 \times 34$  & 0.17 & 0.8 & 3.0 \\
 395 & $2\pi \delta$ & $\pi \delta$ & $54 \times 55 \times 54$ & 0.058 & 0.7 & 2.0\\
 590 & $2\pi \delta$ & $\pi \delta$ & $84 \times 129 \times 94$ & 0.0262 & 0.5 & 4.0\\
\bottomrule
\end{tabular}
    \caption{Main parameters and numerical setup for the simulation of the turbulent channel flow.}
    \label{tab:simulation_param}
\end{table}
\begin{table}[tb!]
    \centering
    \begin{tabular}{ c c c c  }
    \toprule
\multirow{2}{*}{Scheme}  & \multicolumn{3}{c}{$U_b/u_{\tau}$}  \\ 
  & $\Reytau = 180$ & $\Reytau = 395$ & $\Reytau = 590$ \\ 
\midrule
RK3    & {15.65}  & {17.52} & {18.73}   \\
RK4     & {15.52}   & {17.60} &  {18.83}    \\
3p5q(4) & {15.66} & {17.56}  & {18.78}      \\
3p6q(5) & {15.72}   & {17.51} & {19.00}      \\
4p7q(6) & {15.80} & {17.62} & {18.82}    \\ 
\bottomrule
\end{tabular}
    \caption{Ratio between the bulk velocity $U_{b}$ and the friction velocity, for various temporal schemes and $\Reytau = 180$, $395$ and $590$.}
    \label{tab:error_Ub}
\end{table}
The numerical results presented in this section have all been performed with the open-source solver Xcompact3D.
A skew-symmetric form of the convective term is employed, therefore the spatial discretization globally preserves kinetic energy \citep{kravchenko1997effect,verstappen2003symmetry,Coppola2019a}.
Note that the asymmetric schemes used at the wall boundaries in Xcompact3D do not strictly satisfy the summation by parts property, thus there is a non-physical boundary contribution to kinetic energy stemming from the spatial discretization. In order to distinguish this contribution from the one related to the time integration, we explicitly calculated the errors considering the difference between the right-hand side and left-hand side of the discrete global kinetic energy equation \eqref{DiscreteEnergyChannel} below. This difference (of the order $10^{-5}$) is smaller with respect to the temporal error ($10^{-4}$)  investigated and does not affect the presented results.

The channel flow configuration has been investigated for $\Reytau = 180$, $395$, and $590$. Several implicit large-eddy simulations have been performed in a domain of size $4\pi \times 2 \times \frac{4}{3}\pi$ for $\Reytau = 180$ and in a domain of size $2\pi \times 2 \times \pi$ for the other cases. The iLES has been carried out using a spectral vanishing viscosity operator~\citep{Lamballais2011}, by means of $6$th order compact Padé scheme, hence no explicit subgrid model has been used. The main parameters for each simulation, along with the ratio between the hyperviscosity and viscosity $\nu_0/\nu$, which uniquely determine the vanishing viscosity operator~\citep{DAIRAY2017252}, are summarized and reported in Table~\ref{tab:simulation_param}.
 A dedicated analysis of the effect of the accuracy of the spatial scheme is reported in Appendix~\ref{secA1} for the simpler case of the Taylor Green Vortex.
 In addition to the pseudo-symplectic schemes, the  third-order Runge-Kutta-Wray (RK3) and classical fourth-order RK (RK4) schemes have been used to compare the results. The numerical grid is uniform in the homogeneous directions ($x$ and $z$), while the grid in the wall-normal direction is nonuniform and gradually stretched with the use of a tangent hyperbolic function. Therefore, the bilinear forms of the $\varepsilon_\nu$ and $\varepsilon_{\text{RK}}$ have been modified according to the use of a mapping metric term, thus introducing the relevant inner-product to take into account the non-uniform grid in the wall-normal direction \citep{Capuano2018}, see \citep{LAIZET20095990} for more details about the specific metric used. The order of accuracy for the energy conservation has also been verified to be unchanged by the stretching of the mesh. Note that the discrete global kinetic energy equation, Eq.~\eqref{DiscreteEnergyEquation}, has been written for the sake of simplicity in the absence of forcing terms; in this case, a forcing is obviously added to drive the flow through the channel (particularly, a constant pressure gradient is used).

 In such case Eq.~\eqref{DiscreteEnergyEquation} is modified as follows
 \begin{equation}
    \frac{\Delta E}{\Delta t} = \frac{1}{Re}\sum_{i=1}^{s} b_j \vec{{u_j}}^T  \mtx{L}\vec{u_j} - \frac{\Delta t}{2} \sum_{i,j=1}^{s} \left(b_i a_{ij} + b_j a_{ji}-b_j b_i\right) \mtx{{\Tilde{f}_i}}^T\mtx{\Tilde{f}_j} + q_1 \sum_{i=1}^{s} b_i\vec{u_i}^T \vec{e_x} 
    \label{DiscreteEnergyChannel}
\end{equation}
where $q_1$ represents the forcing term in the channel flow and $\vec{e_x} = [\vec{e_1},\vec{0},\vec{0}]$, $\vec{e_1}, \vec{0} \in \mathbb{R}^{{N_{\boldsymbol{u}}/3}}$. Therefore, Eq.~\eqref{DiscreteEnergyEquation} can also be written equivalently as follows
 \begin{equation}
    \frac{\Delta E}{\Delta t} = \frac{1}{Re} \Phi + \varepsilon_{\text{RK}} + \varepsilon_{\text{CPG}} = \frac{1}{\Reyeff} {\Phi} + \varepsilon_{\text{CPG}}
\end{equation}
where 
\begin{equation}
    \varepsilon_{\text{CPG}} = q_1 \sum_{i=1}^{s} \vec{u_i}^T \vec{e_x} 
\end{equation}
represents the discrete counterpart of the of the forcing term appearing in the continuous energy equation, thus the definition of the effective Reynolds number has not been modified, since the forcing term does not result in any additional terms that could be classified as temporal errors.
 The $\CFL$ number reported in Table \ref{tab:simulation_param} is defined with respect to the initial condition as $\CFL = \max\left(\max_{{ijk}}\left(\frac{|u_{ijk}| \Delta t}{\Delta x}, \frac{|v_{ijk}| \Delta t}{\Delta y},\frac{|w_{ijk}| \Delta t}{\Delta z} \right) \right)$.
 For sake of completeness the $\CFL$ numbers once the channel is fully developed are $\CFL = 0.68$, $0.7$ and $0.73$ respectively for $\Reytau = 180$, $395$ and $590$. Importantly, for each case, in order to reduce the sampling error and collect temporally-independent spatial fields, the samples have been spaced of about $0.1 \tau$ in time, over a time window of about $10 \tau$, where $\tau = h/u_{\tau}$~\citep{Pirozzoli}. Moreover, accompanying simulations have also been carried out, computed at $\CFL = 0.01$. These are considered to be virtually free of temporal errors, and are therefore regarded as \textit{reference} results.

\subsection{Global behaviour of the temporal error}
\begin{figure*}[tb!]
    \centering
    \begin{subfigure}[b]{0.32\linewidth}
        \includegraphics*[width = \textwidth]{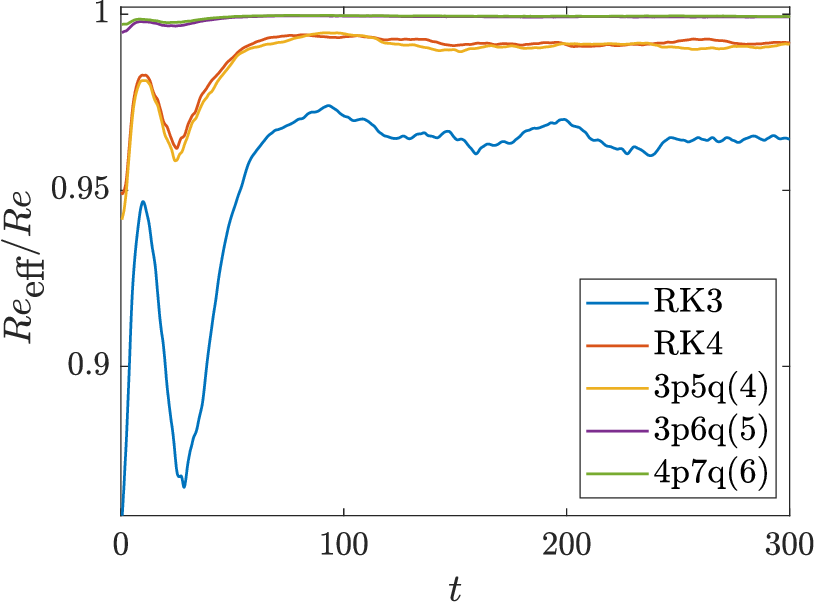}
        \caption{CFL = 0.8, $\Reytau = 180$.}
        \label{fig:Reynolds_time180}
    \end{subfigure}
    \begin{subfigure}[b]{0.32\linewidth}
        \includegraphics*[width = \textwidth]{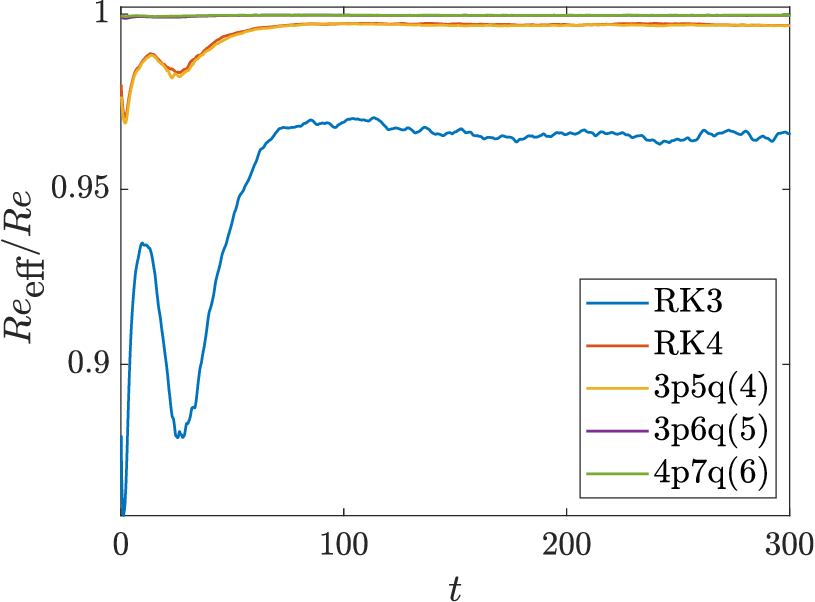}
        \caption{CFL = 0.7, $\Reytau = 395$.}
        \label{fig:Reynolds_time395}
    \end{subfigure}
    \begin{subfigure}[b]{0.32\linewidth}
        \includegraphics*[width = \textwidth]{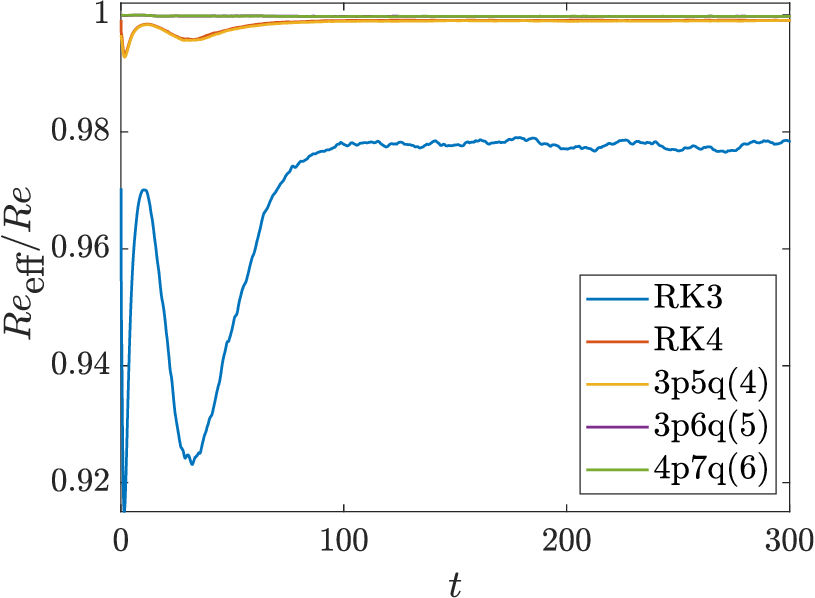}
        \caption{CFL = 0.5, $\Reytau = 590$.}
        \label{fig:Reynolds_time590}
    \end{subfigure}
     \caption{Ratio of effective to nominal Reynolds number for the numerical simulation of the turbulent channel flow.}
    \label{fig:Reynolds_effective_channel}
\end{figure*}

In  Figures~\ref{fig:Reynolds_time180},~\ref{fig:Reynolds_time395} and~\ref{fig:Reynolds_time590} the ratio of the effective Reynolds number with respect to the nominal Reynolds number for $\Reytau = 180$, $395$ and $590$ are shown as a function of time. The second drop corresponds to the starting point of the transition to turbulence, until it reaches a condition where the ratio can be considered to be stationary; hence the channel flow has reached the condition of being fully developed.
The figures do not show noticeable differences between the pseudo-symplectic schemes of higher order, i.e.,~4p7q(6) and 3p6q(5), while on the other hand RK3, RK4 and 3p5q(4) schemes vary significantly, reaching a deviation from the nominal Reynolds number up to 4\% for the RK3 and up to 1\% for the four-stage RK schemes at $\Reytau = 395$ after the transient phase. The similar behaviour of the fourth stage RK methods, namely RK4 and 3p5q(4), can be traced back to the pseudo-symplectic conditions of the RK4 method for 5th-order accuracy, where only one of these conditions is not satisfied. Therefore, in some cases (and for some flows), this term can be so small that the RK4 achieves 5th-order accuracy on energy.  
In Table~\ref{tab:error_Ub} the bulk velocity $U_b$ normalized with the friction velocity for different temporal schemes is reported. The pseudo-symplectic schemes of higher-order are able to keep the lowest level of production of artificial dissipation, with a maximum error below 0.1\% after the transition phase.

\subsection{First- and second-order statistics}
\begin{figure*}[tb!]
    \centering
    \begin{subfigure}[b]{0.32\linewidth}
        \includegraphics*[width = \textwidth]{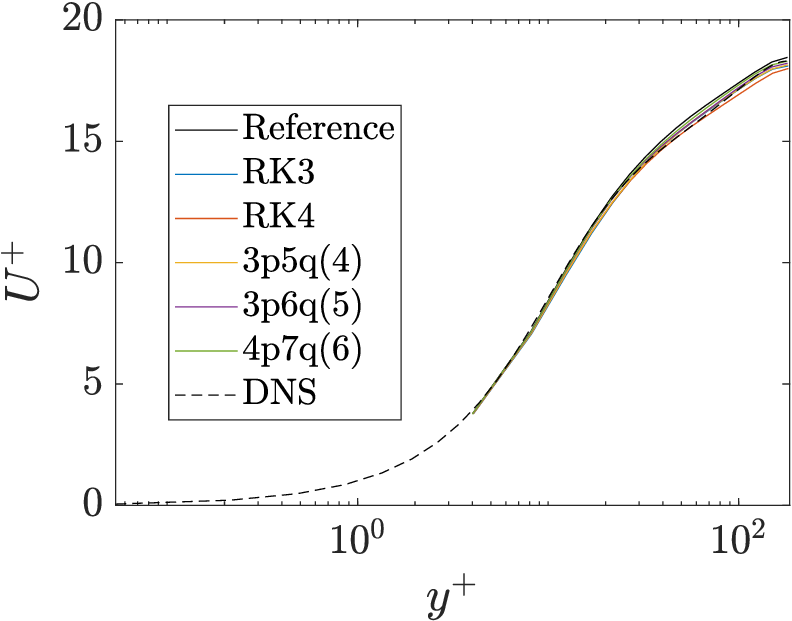}
        \caption{CFL = 0.8, $\Reytau = 180$.}
        \label{fig:Reynolds_umean180}
    \end{subfigure}
    \begin{subfigure}[b]{0.32\linewidth}
        \includegraphics*[width = \textwidth]{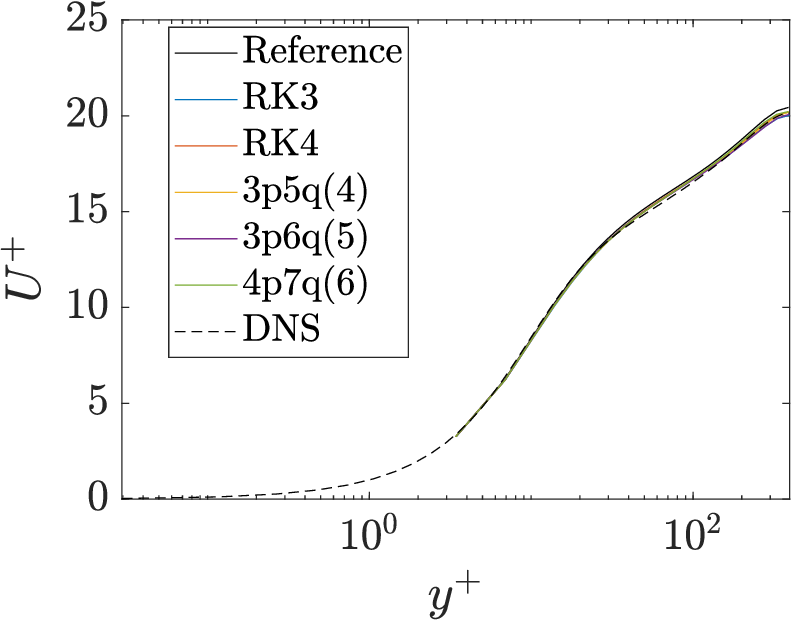}
        \caption{CFL = 0.7, $\Reytau = 395$.}
        \label{fig:Reynolds_umean395}
    \end{subfigure}
    \begin{subfigure}[b]{0.32\linewidth}
        \includegraphics*[width = \textwidth]{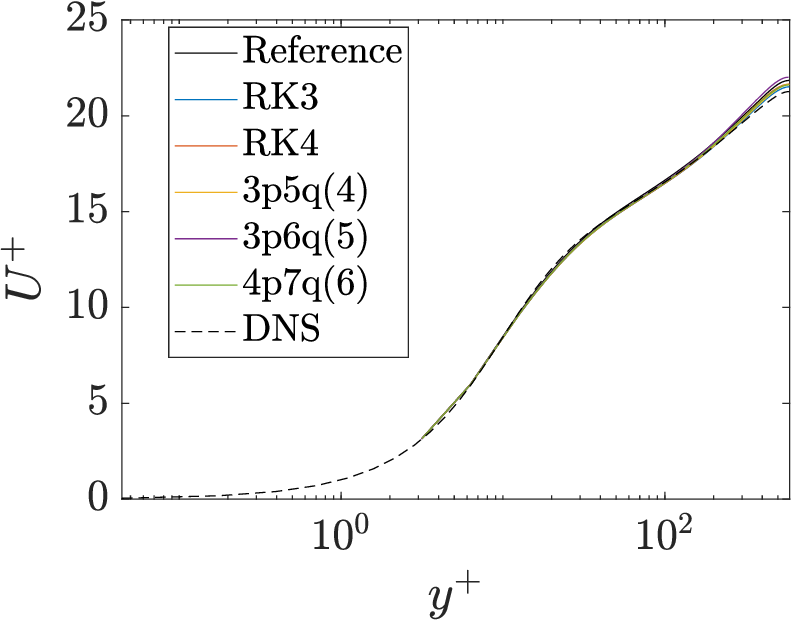}
        \caption{CFL = 0.5, $\Reytau = 590$.}
        \label{fig:Reynolds_umean590}
    \end{subfigure}
     \caption{Mean streamwise velocity component for the numerical simulation of the turbulent channel flow with respect to a reference solution computed at CFL = 0.01 and DNS reference data~\citep{MoserDNSreferencedata}.}
    \label{fig:Reynolds_umean_channel}
\end{figure*}
\begin{figure*}[tb!]
    \centering
    \begin{subfigure}[b]{0.32\linewidth}
        \includegraphics*[width = \textwidth]{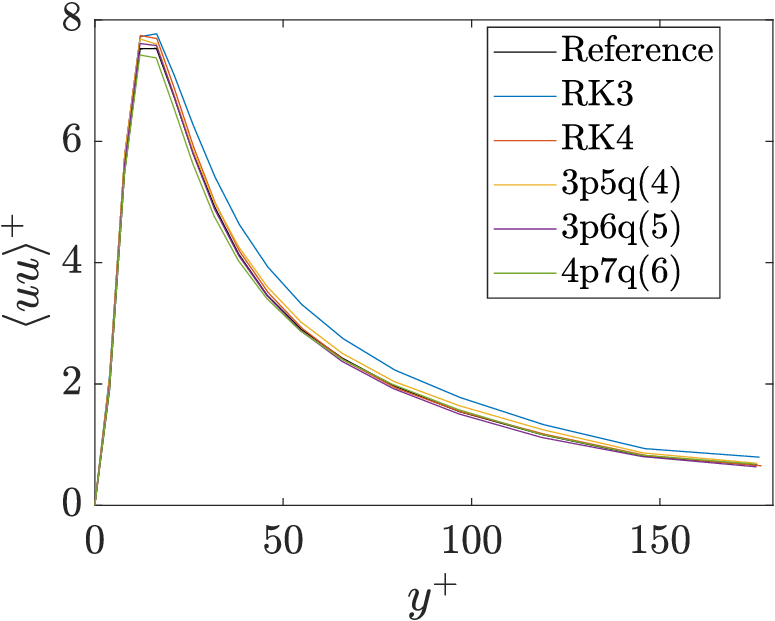}
        \caption{CFL = 0.8, $\Reytau = 180$.}
        \label{fig:Reynolds_uumean180}
    \end{subfigure}
    \begin{subfigure}[b]{0.32\linewidth}
        \includegraphics*[width = \textwidth]{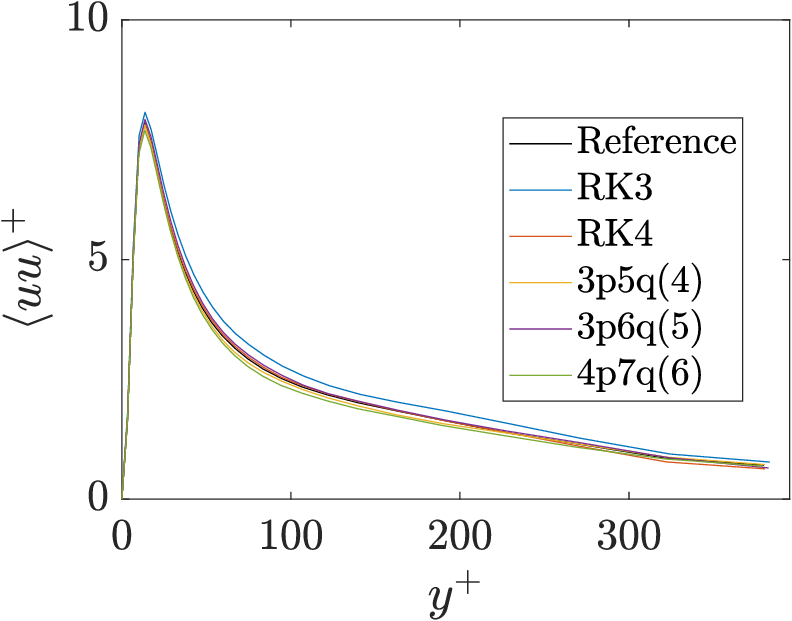}
        \caption{CFL = 0.7, $\Reytau = 395$.}
        \label{fig:Reynolds_uumean395}
    \end{subfigure}
    \begin{subfigure}[b]{0.32\linewidth}
        \includegraphics*[width = \textwidth]{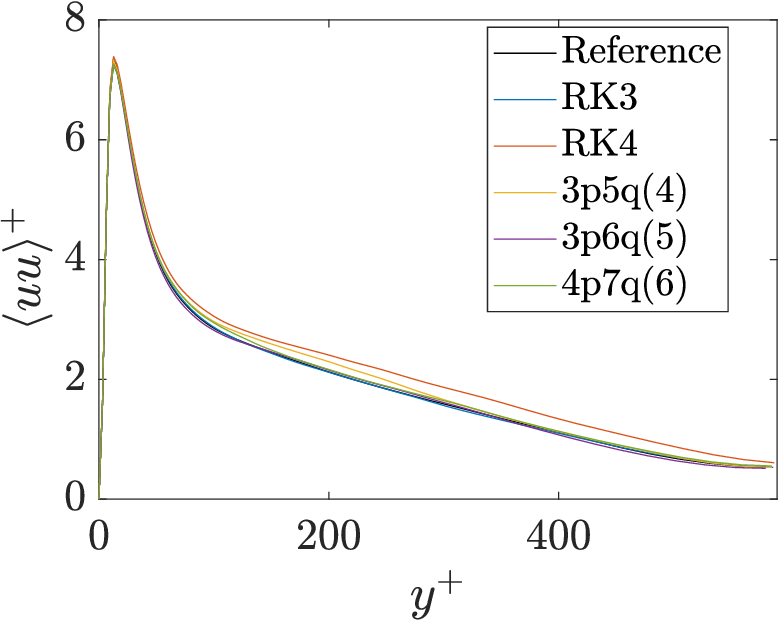}
        \caption{CFL = 0.5, $\Reytau = 590$.}
        \label{fig:Reynolds_uumean590}
    \end{subfigure}
     \caption{Variance of $u$ for the numerical simulation of the turbulent channel flow with respect to a reference solution computed at CFL = 0.01.}
    \label{fig:Reynolds_uumean_channel}
\end{figure*}
\begin{figure*}[tb!]
    \centering
    \begin{subfigure}[b]{0.32\linewidth}
        \includegraphics*[width = \textwidth]{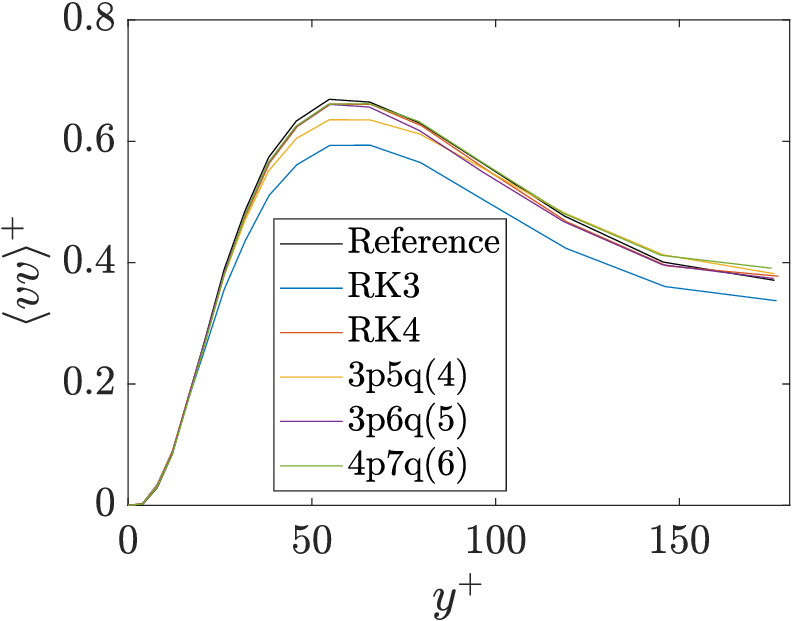}
        \caption{CFL = 0.8, $\Reytau = 180$.}
        \label{fig:Reynolds_vvmean180}
    \end{subfigure}
    \begin{subfigure}[b]{0.32\linewidth}
        \includegraphics*[width = \textwidth]{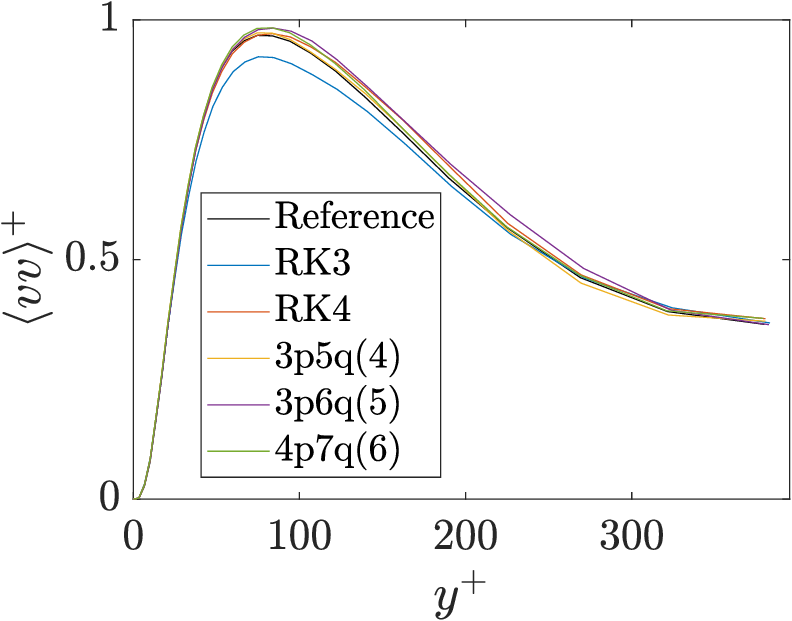}
        \caption{CFL = 0.7, $\Reytau = 395$.}
        \label{fig:Reynolds_vvmean395}
    \end{subfigure}
    \begin{subfigure}[b]{0.32\linewidth}
        \includegraphics*[width = \textwidth]{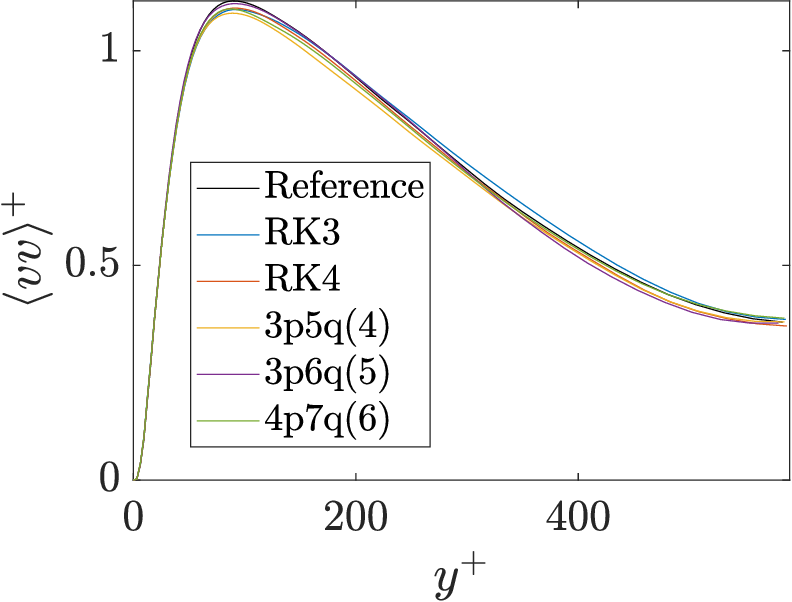}
        \caption{CFL = 0.5, $\Reytau = 590$.}
        \label{fig:Reynolds_vvmean590}
    \end{subfigure}
     \caption{Variance of $v$ for the numerical simulation of the turbulent channel flow with respect to a reference solution computed at CFL = 0.01.}
    \label{fig:Reynolds_vvmean_channel}
\end{figure*}
\begin{figure*}[tb!]
    \centering
    \begin{subfigure}[b]{0.32\linewidth}
        \includegraphics*[width = \textwidth]{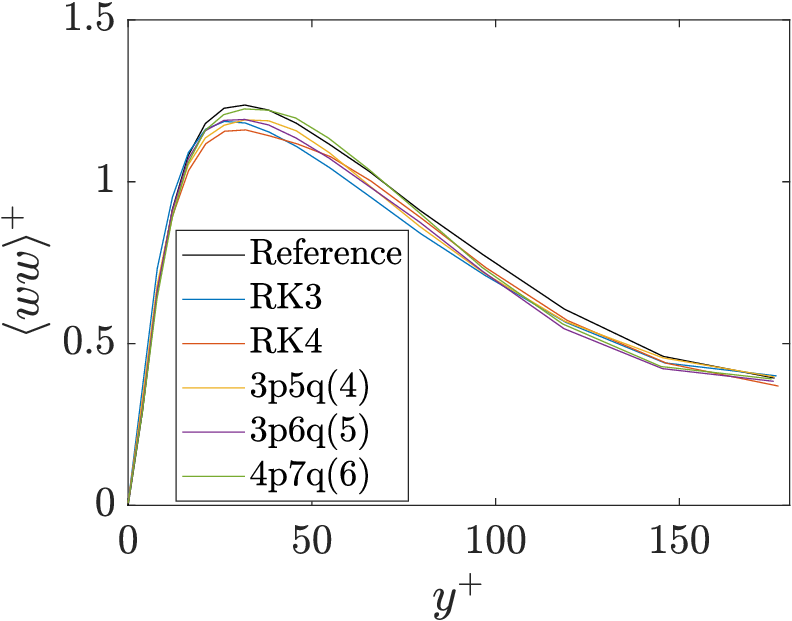}
        \caption{CFL = 0.8, $\Reytau = 180$.}
        \label{fig:Reynolds_wwmean180}
    \end{subfigure}
    \begin{subfigure}[b]{0.32\linewidth}
        \includegraphics*[width = \textwidth]{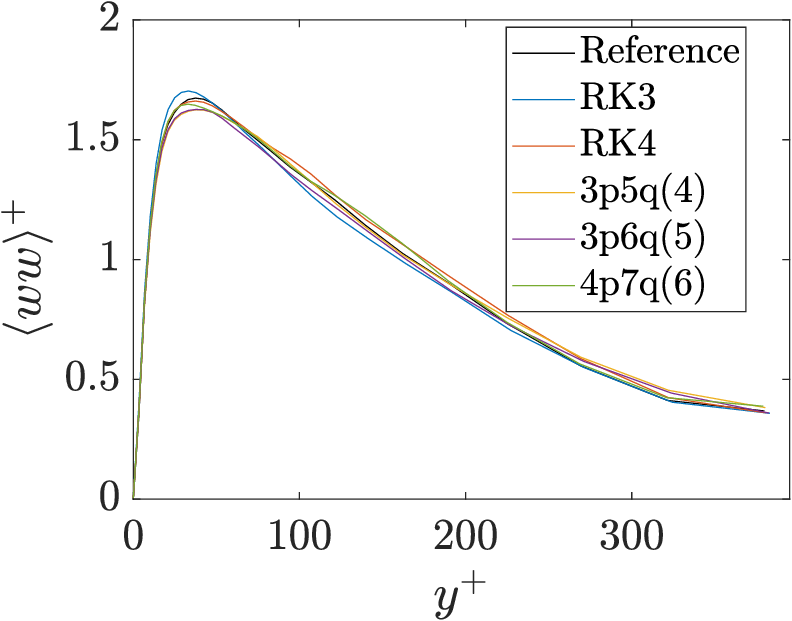}
        \caption{CFL = 0.7, $\Reytau = 395$.}
        \label{fig:Reynolds_wwmean395}
    \end{subfigure}
    \begin{subfigure}[b]{0.32\linewidth}
        \includegraphics*[width = \textwidth]{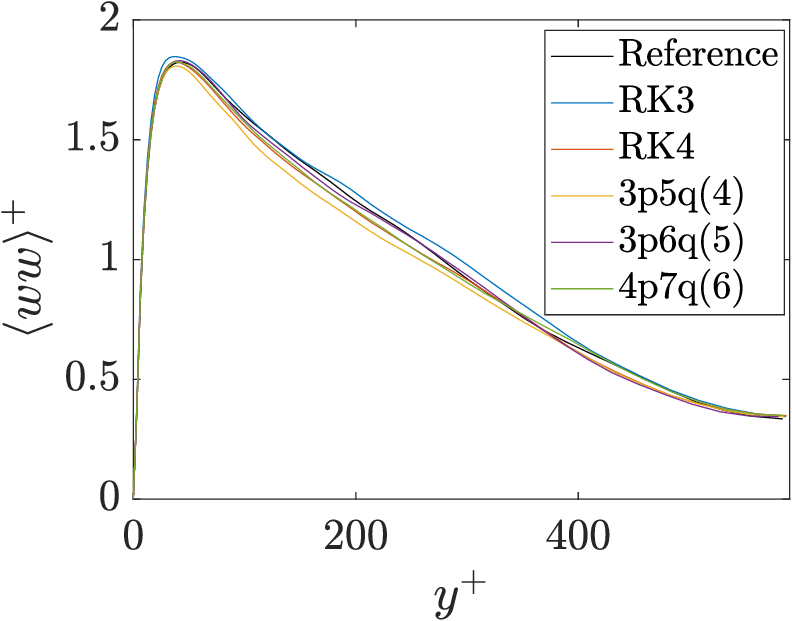}
        \caption{CFL = 0.5, $\Reytau = 590$.}
        \label{fig:Reynolds_wwmean590}
    \end{subfigure}
     \caption{Variance of $w$ for the numerical simulation of the turbulent channel flow with respect to a reference solution computed at CFL = 0.01.}
    \label{fig:Reynolds_wwmean_channel}
\end{figure*}

The first- and second-order statistics of wall-normal profiles are reported in Figure~\ref{fig:Reynolds_umean_channel}.
Some discrepancies can be highlighted for the components of the Reynolds stress tensor, with an error that results generally higher compared with first-order statistics. In particular, for the variance of the vertical velocity component the differences are the most noticeable, as observed in Figures~\ref{fig:Reynolds_vvmean180} and \ref{fig:Reynolds_vvmean395}.
The variance for the other velocity components is reported in Figures~\ref{fig:Reynolds_uumean_channel}
and~\ref{fig:Reynolds_wwmean_channel}.
It is important to highlight that the commonly used RK3 significantly underestimates the peak of the wall-normal stress profile at CFL$=0.8$, which lies in the medium-upper range of commonly used Courant numbers for large-eddy simulation of turbulent flows.

\subsection{Energy Spectra}
The energy spectrum is a powerful mean to represent the dissipative behaviour of turbulent flows. The use of an over-dissipative scheme by means of a spectral vanishing viscosity operator has been demonstrated to efficiently compensate the small-scale dissipation 
that is missed by
a coarse grid~\citep{inbook}. Here we focused on the effect of time integration scheme on the turbulent energy spectra for large-eddy simulations of a turbulent channel flow. Time-averaged energy spectra in the streamwise direction have been computed at the centerline of the channel.  Moreover, a reference solution has been computed at a small $\Delta t$ ($\CFL = 0.01$) to compare the different behaviour of the temporal schemes.

\begin{figure}[htb!]
	\centering
    \begin{subfigure}[b]{0.48\linewidth}
	\includegraphics*[width=0.98\linewidth]{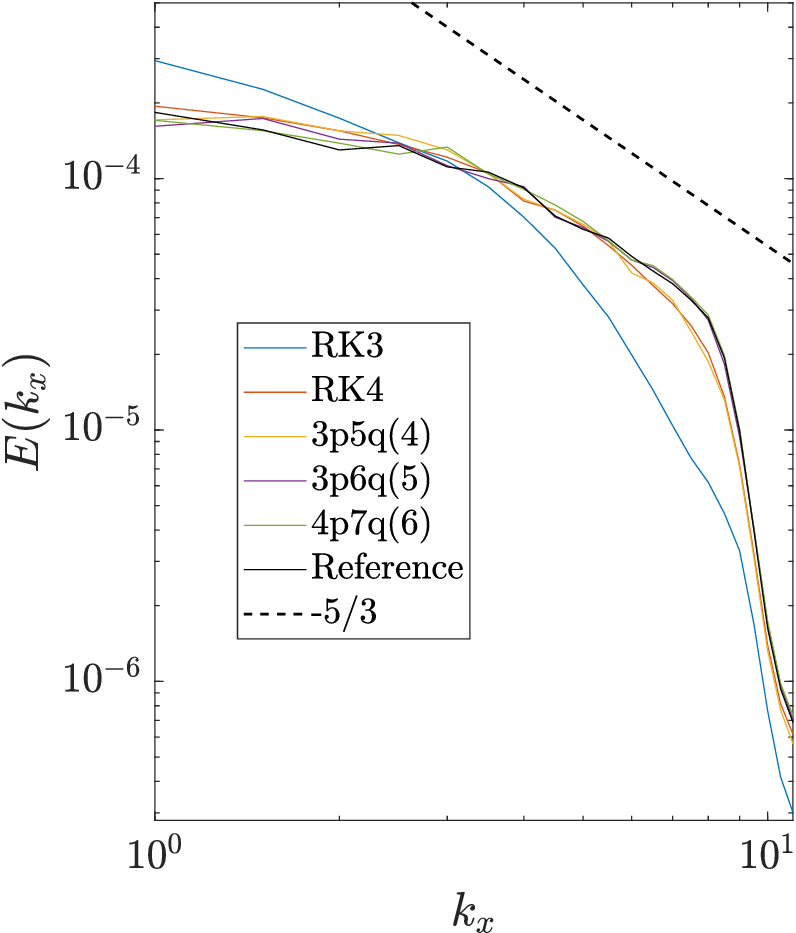}
	\caption{CFL = 0.8, $\Reytau = 180$.}
 \label{ES180}
	\end{subfigure} \hspace{0.2cm}
    \begin{subfigure}[b]{0.48\linewidth}
	\includegraphics*[width=0.98\linewidth]{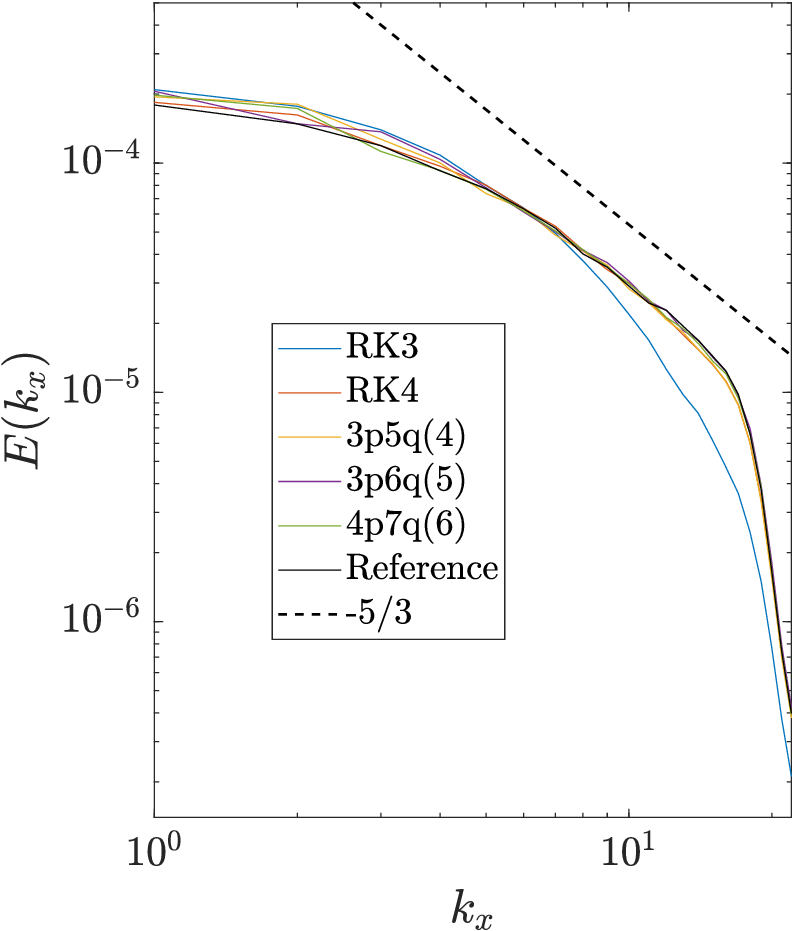}
	\caption{CFL = 0.7, $\Reytau = 395$.}
 \label{ES395}
  \end{subfigure}
  \\
\begin{subfigure}[b]{0.48\linewidth}
	\includegraphics*[width=0.98\linewidth]{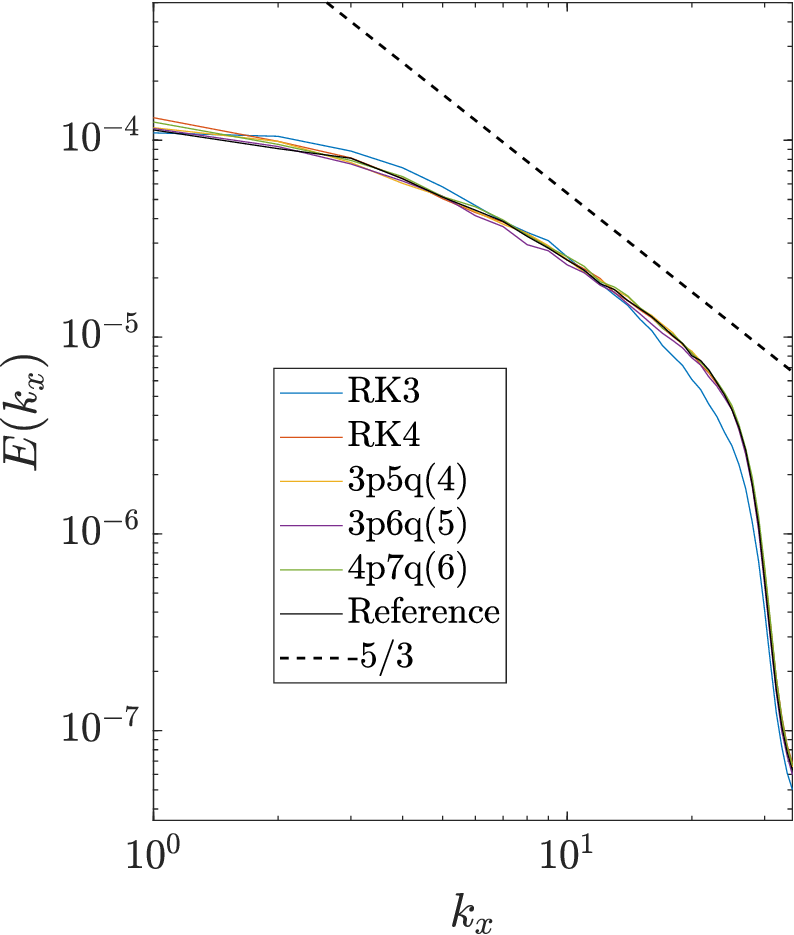}
	\caption{CFL = 0.5, $\Reytau = 590$.}
 \label{ES590}
	\end{subfigure}
 \caption{Time-averaged streamwise energy spectra at the centerline for different Reynolds numbers.}
\end{figure}

Figure~\ref{ES180} illustrates that at small wavenumbers, RK3 exhibits a significant increase in energy for $\Reytau = 180$. This behaviour is similarly observed in the energy spectra for $\Reytau = 395$ and $590$, as shown in Figures~\ref{ES395} and~\ref{ES590}. As the large eddies break up and supply energy to the smaller scales, the energy spectra of RK3, RK4 and 3p5q(4) show a significant variation for $\Reytau = 180$. Indeed the higher dissipative error in the conservation of the discrete kinetic energy is responsible for the significant damping in the energy spectrum for the RK3 scheme towards marginally-resolved wavenumbers. However, as the Reynolds number increases, these differences become less pronounced. Thus, the choice of time-integration scheme substantially influences the energy transfer mechanism, resulting in a higher dissipation rate in the inertial range, thereby compromising the physical realism of the solution. This behaviour extends also down to the smallest scales, consequently reducing the value of the cutoff wavenumber of the LES. On the other hand, the high-order pseudo-symplectic schemes preserve the dynamics of the physical energy cascade across all the investigated $\Reytau$, exhibiting a good agreement with the reference solution.

\subsection{Performance Analysis}
\begin{figure}[tb!]
	\begin{center}
	\includegraphics*[width=0.6\linewidth]{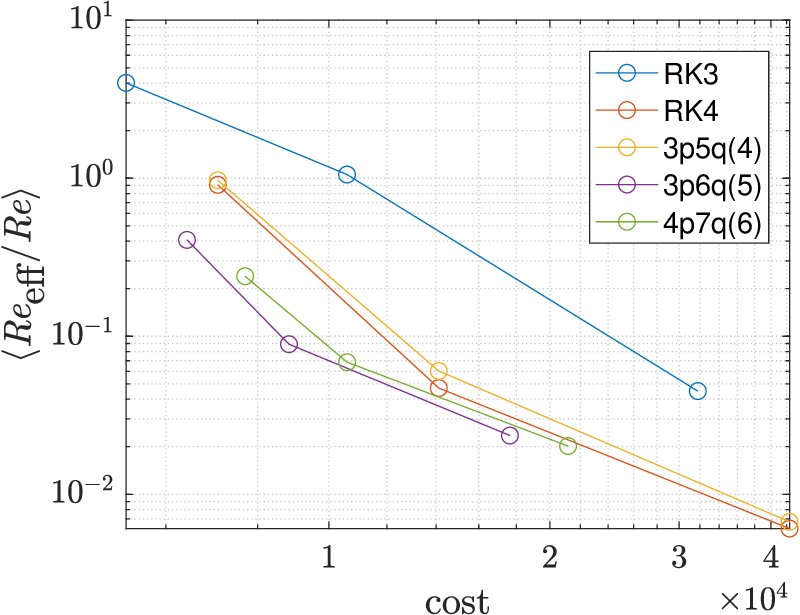}
	\caption{\label{fig:Reynolds_cost180} Averaged ratio of effective to nominal Reynolds number for the LES as a function of number of right-hand side evaluations for $\Reytau = 180$ and CFL varying from $0.1$ to $0.8$.}
	\end{center}
\end{figure}

\begin{figure}[tb!]
	\begin{center}
	\includegraphics*[width=0.6\linewidth]{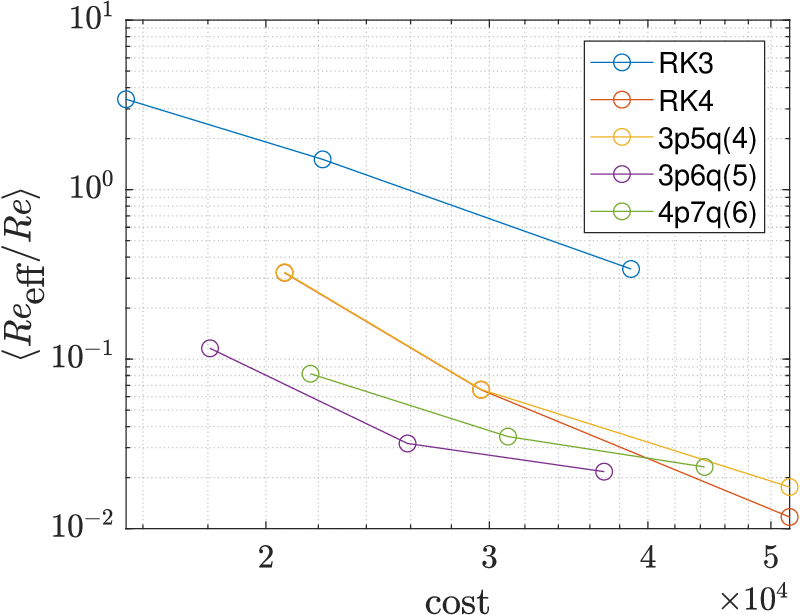}
	\caption{\label{fig:Reynolds_cost395}Averaged ratio of effective to nominal Reynolds number for the LES as a function of number of right-hand side evaluations for $\Reytau = 395$ and CFL varying from $0.2$ to $0.7$.}
	\end{center}
\end{figure}

\begin{figure}[tb!]
	\begin{center}
	\includegraphics*[width=0.6\linewidth]{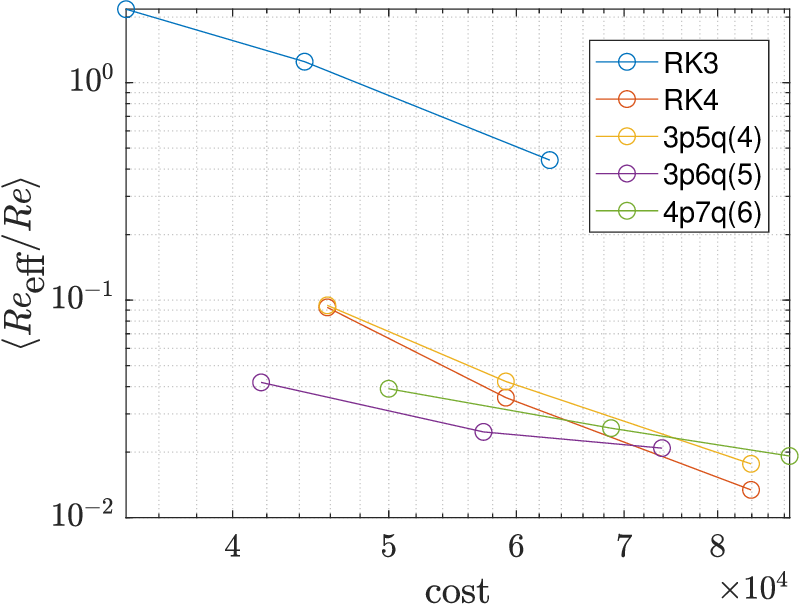}
	\caption{\label{fig:Reynolds_cost590}Averaged ratio of effective to nominal Reynolds number for the LES as a function of number of right-hand side evaluations for $\Reytau = 590$ and CFL varying from $0.2$ to $0.5$.}
	\end{center}
\end{figure}

A meaningful comparison between the RK schemes can be achieved by means of a cost-vs-accuracy analysis. Here, the averaged ratio of effective to nominal Reynolds number at the statistical steady state is chosen as the accuracy measure and reported together with a cost function, which is defined as the number of the right-hand side evaluations required to reach the fully developed channel flow with different time step sizes. As it is shown in Figures~\ref{fig:Reynolds_cost180},~\ref{fig:Reynolds_cost395} and~\ref{fig:Reynolds_cost590}, the analysis shows that higher order methods, in particular 3p6q(5) and 4p7q(6), are the most cost-effective and efficient among the temporal schemes investigated, as they require the minimum cost for a given value of the error. However, the 3p6q(5) performs slightly better than the more accurate pseudo-symplectic scheme of higher-order 4p7q(6), as already shown for different boundary and initial conditions and numerical setup by \cite{Capuano2019}, which is confirmed within the range of $\Reytau$ investigated. 
The better performances of higher-order pseudo-symplectic schemes are evident for a wide range of \CFL~numbers, corresponding to low to moderate computational costs, while the benefits become less pronounced for very small time step sizes.

\section{Conclusions}\label{sec:conclusions}
The temporal error of standard and pseudo-symplectic RK methods has been investigated for incompressible flows. 
The improved energy-conservation properties of pseudo-symplectic schemes have been assessed also for a turbulent channel flow, being able to keep the error on the preservation of global kinetic energy below 1\%, and minimizing the temporal error.
Additionally, different temporal integration schemes revealed significant effects on the energy dissipation and the fidelity of turbulent energy spectra representation. The use of high-order pseudo-symplectic schemes proved critical in preserving the dynamics of the physical energy cascade, showing good agreements with the reference solution. Conversely, low-order schemes, such as the RK3 by \cite{Wray1991} can lead to a loss of physical realism in spectral representation, particularly in the inertial range up to the smallest scales, reducing the value of the cutoff wave number for the LES. Thus, the analysis suggests that the choice of temporal integration plays a crucial role in accurately representing subtle turbulent phenomena in wall-bounded flow large-eddy simulations. 
Furthermore, a cost-vs-accuracy analysis for the channel flow results shows that high-order pseudo-symplectic schemes, such as the 3p6q(5) method, are the most efficient and cost-effective ones among other pseudo-symplectic schemes and standard RK methods.

\appendix

\section{Effect of spatial scheme}\label{secA1}

To determine the influence of the spatial schemes on the effective Reynolds number (which is indicative of the temporal kinetic-energy preservation error), coarse direct numerical simulatins (DNS) of the Taylor-Green-Vortex at $\Rey = 3000$ and $\CFL = 1$ have been performed in a tri-periodic square domain of side length $L = 2\pi$, discretized using $N = 65^3$ nodes. In particular, the temporal error of classical and pseudo-symplectic RK schemes have been compared by employing different spatial schemes: central $2$nd order, compact $4$th and $6$th order Padé schemes.

\begin{figure}[tb!]
	\begin{center}
	\includegraphics[width=0.5\linewidth]{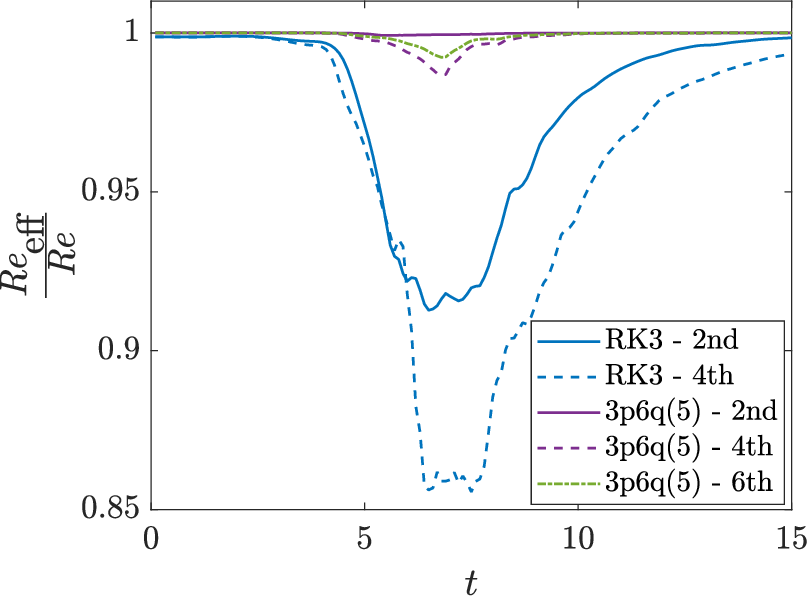}
	\caption{Comparison of the effective Reynolds number for the Taylor-Green-Vortex at $\Rey = 3000$, $N = 65^3$, for different temporal and spatial schemes, at $\CFL = 1$.}\label{fig:EffReTGV}
	\end{center}
\end{figure}

\begin{figure*}[tb!]
    \centering
    \begin{subfigure}[b]{0.32\linewidth}
        \includegraphics*[width=\linewidth]{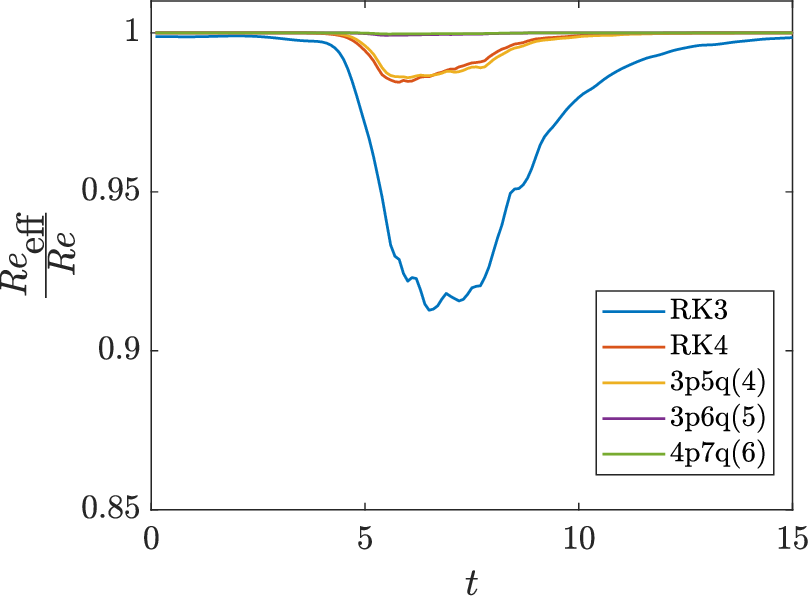}
        \caption{Central second order}
        \label{fig:Reynolds_eff_TGV_2}
    \end{subfigure}
    \begin{subfigure}[b]{0.32\linewidth}
        \includegraphics*[width=\linewidth]{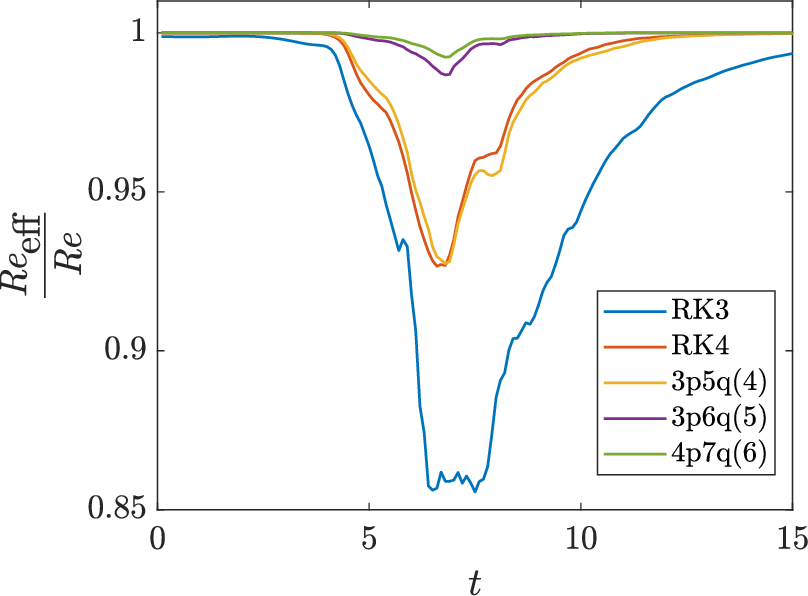}
        \caption{Compact fourth order}
        \label{fig:Reynolds_eff_TGV_4}
    \end{subfigure}
    \begin{subfigure}[b]{0.32\linewidth}
        \includegraphics*[width=\linewidth]{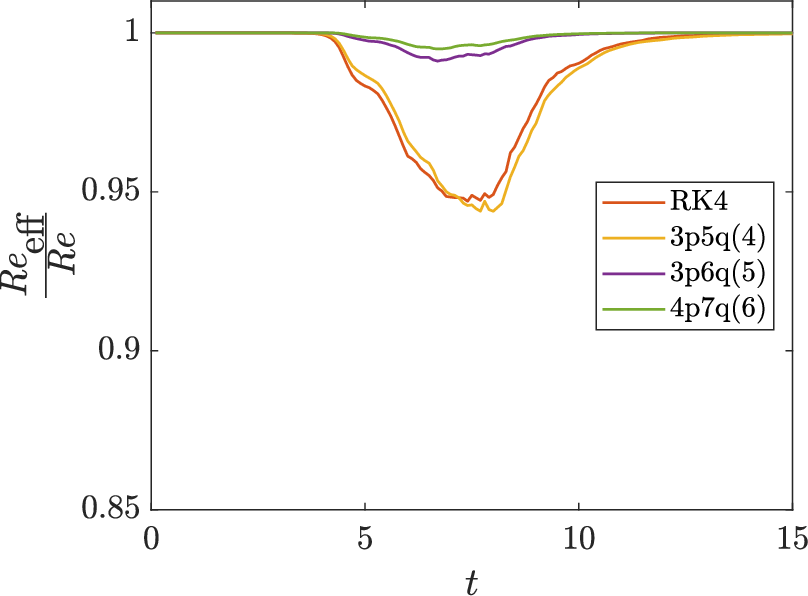}
        \caption{Compact sixth order}
        \label{fig:Reynolds_eff_TGV_6}
    \end{subfigure}

     \caption{Ratio of effective to nominal Reynolds number at $\CFL = 1$ for the Taylor-Green-Vortex at $\Rey = 3000$, $N = 65^3$ with different spatial schemes.}
    \label{fig:Reynolds_effective_TGV_Full}
\end{figure*}

In Figure~\ref{fig:EffReTGV} the time evolution of the effective Reynolds number $\Reyeff/\Rey = \frac{\varepsilon_\nu}{\varepsilon_{\text{RK}} + \varepsilon_\nu}$ is reported, in which $\varepsilon_\nu = \Phi/\Rey$ is the physical dissipation rate.
For both RK3 and 3p6q(5), the use of the higher order spatial schemes leads to a higher production of artificial dissipation than the central second order spatial scheme. The origin of this behaviour has been investigated by studying the physical dissipation rate $\varepsilon_\nu$ and temporal error $\varepsilon_{\text{RK}}$ independently.
The influence of the spatial discretization order on $\varepsilon_\nu$ is much smaller than the effect on $\varepsilon_{\text{RK}}$, so it is the last one that is responsible for the noticeable change in the effective Reynolds number.
The same behaviour has been found to hold also for the other temporal integrators tested, with compact $4th$ compact schemes leading to the highest artificial viscosity among all the spatial schemes.
In Figure~\ref{fig:Reynolds_effective_TGV_Full} the effective Reynolds number for the central-second order, compact fourth order and sixth order are reported. The RK3 with compact sixth order in Figure~\ref{fig:Reynolds_eff_TGV_6} has not been reported due to loss of numerical stability for the high CFL used.


\subsubsection*{Acknowledgments}

Part of this work was supported by a grant of HPC time from CINECA under the ISCRA project \emph{PSRK}. Francesco Capuano is a Serra H\'unter fellow. Marco Artiano was supported by the Daimler und Benz Stiftung (Daimler and Benz foundation, project number 32-10/22).
\subsubsection*{Author Contributions}
M.~A.: Conceptualization, Methodology, Investigation, Writing – original draft, Writing – review \& editing.
C.~D.~M.: Conceptualization, Methodology, Investigation, Writing – original draft, Writing – review \& editing.
F.~C.: Conceptualization, Methodology, Investigation, Writing – original draft, Writing – review \& editing, Supervision.
G.~C.: Conceptualization, Methodology, Investigation, Writing – original draft, Writing – review \& editing, Supervision.
\subsubsection*{Funding}
Marco Artiano was supported by the Daimler und Benz Stiftung (Daimler and Benz foundation, project number 32-10/22).
\subsubsection*{Data Availability}
Data can be made available upon request to the corresponding author.

\section*{Declarations}
\subsubsection*{Conflict of Interest}
The authors have no relevant financial or non-financial interests to disclose.
\subsubsection*{Ethical approval}
Not applicable.
\subsubsection*{Informed consent}
Not applicable.
\bibliographystyle{abbrvnat}
\bibliography{refs} 
\end{document}